\documentclass[intlimits,twoside,a4paper]{article}

\usepackage{amsmath,amssymb}
\usepackage{graphicx}

\usepackage[T2A]{fontenc}
\usepackage[cp1251]{inputenc}

\usepackage[eqsecnum]{cmpj2}
\issue{2016}{19}{3}{33401}
\doinumber{10.5488/CMP.19.33401}

\title[Local field enhancement at the core of cylindrical nanoinclusions]%
{Local field enhancement at the core of cylindrical nanoinclusions  embedded in a linear dielectric host matrix%
}
\author[Y.A.~Abbo, V.N.~Mal'nev, A.A.~Ismail]{Y.A.~Abbo\refaddr{label1,label2}\thanks{E-mail: yilakaa@yahoo.com}\,, V.N.~Mal'nev\refaddr{label2}, A.A.~Ismail\refaddr{label2}}
\addresses{
\addr{label1}Addis Ababa (Asko), P.O.BOX 171078, Ethiopia
\addr{label2} Department of Physics, Addis Ababa University, P.O.BOX 1176, Addis Ababa, Ethiopia
}

\authorcopyright{Y.A.~Abbo, V.N.~Mal'nev, A.A.~Ismail, 2016}
\date{Received October 11, 2015, in final form March 16, 2016}
\begin{document}

\maketitle

\begin{abstract}
 In this paper we have discussed theoretical concepts  and presented numerical results of local field enhancement at the core of different assemblages of metal/dielectric  cylindrical nanoinclusions embedded in a linear dielectric host matrix. The obtained results show that for a composite with metal coated inclusions there exist two peak values of the enhancement factor at two different resonant frequencies. The existence of the second maxima becomes more important for a larger volume fraction of the metal part of the inclusion. For dielectric coated metal core inclusions and pure metal  inclusions there is only one resonant frequency and one peak value  of the enhancement factor. The enhancement of an electromagnetic wave is promising for the existence of nonlinear optical phenomena such as optical bistability which  is important in optical communication and in optical computing such as  optical switch and memory elements.
\pacs 42.65.Pc, 42.79.Ta, 78.67.Sc, 78.67.-n
\end{abstract}
\section{Introduction}

Classifying the materials based on their electrical, optical, magnetic or chemical properties in the  scientific community embraces a longer period of time. In another way, the  materials can be sorted based on the number of their constituents. According to the number of their constituents, the materials can be classified as monolithic or composite. A monolithic material has a single constituent, while composite materials are made of two or more constituent materials having significantly different physical and chemical properties, that, when combined, produce a material with a characteristic different from the individual components. In a composite material, one of the constituents is a continuous matrix which is called a host matrix while the others dispersed in the host matrix are called inclusions or fillers \cite{Chen04, Gre04, url1}.

In recent years, the materials with nano-sized inclusions become an active area of research because they have amusing optical properties and applications without significant  physical change in the bulk host medium. Such a material with nano-sized dispersed particles in the matrix where the dimensions of the particles are within the scale of nanometer at least in one direction known as nanocomposites.

The properties of a composite material are related to the properties and fraction of the constituents, so the electromagnetic properties of the composites can be tailored by varying the properties and fractions of the constituents \cite{url1}. Composite materials, consisting of small nonlinear metallic particles (existing in the shape of ellipsoid, sphere or cylinder) are preferred due to their complex responses to the incident light field. Composites are of particular interest because their nonlinearity may be strongly enhanced relative to the bulk sample of the same materials \cite{Pan01}, Amusing nonlinear phenomena, for example optical bistability, have been predicted theoretically \cite{Len86} and have been observed experimentally \cite{Ber94}. Their promising applications include optical switches in optical communication and in optical computing, as well as in transistor, pulse shaper, and memory elements \cite{Len86, Rob07, Gao04}.

The electric field within an atom has the magnitude of $E_{\text{at}}=5.14 \times 10^{11}$~V/m  and the laser intensity associated with a peak field strength of $ E_{\text{at}}$ is $ I_{\text{at}}= 3.5 \times 10^{16}$~W/cm$^{2} $. When the applied electric field exceeds the atomic electric field strength, nonlinear optical phenomenon such as optical bistability may occur. The term optical bistability refers to the situation in which two different output intensities are possible for a given input intensity \cite{Rob07}.

It may be insufficient to apply an ordinary intense electromagnetic wave such as laser in order to achieve an optical nonlinearity. Thus, the applied field should be enhanced. One way of enhancing the applied electromagnetic field is to modify the property of the bulk medium by introducing very small inclusions, such as nanoinclusions of different shapes.

O.A.~Buryl  et al. studied the local field enhancement at the core of elliptical nanoinclusions in a dielectric host matrix \cite{Bur11} and, Sisay and Mal'nev  studied the local field enhancement at the core of spherical  nanoinclusions in a linear dielectric host matrix \cite{Sis13}. The results of these investigations surpass the results of abnormal enhancement of the local field, when the frequency of the incident electromagnetic wave approaches the surface plasmon frequency of the metal part of the inclusions.

To our knowledge, the existence of the second peak value of the enhancement factor was first presented by Sisay and Mal'nev \cite{Sis13} for a composite with metal coted spherical nanoinclusions. In this paper, we have investigated the local field enhancement at the core of cylindrical metal/dielectric nanoinclusions embedded in a linear dielectric host matrix. While studying the results of early works, we considered that in addition to different physical quantities, the shape of the inclusion also plays a great role in varying the magnitude of the local field enhancement factor at the core of the inclusions. This work  confirms the existence of the second peak value of the enhancement factor for a composite with metal coated dielectric core cylindrical nanoinclusion. To our knowledge, this original work  shows the existence of the second maxima of the enhancement factor for a composite having cylindrical inclusions.

\section{Electrical potential distribution}

\begin{figure}[!t]
\centerline{\includegraphics[width=0.55\textwidth]{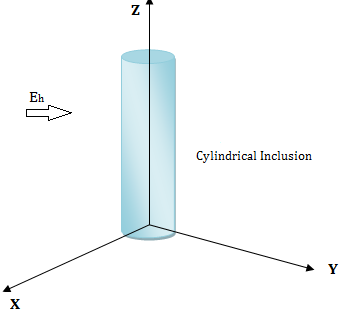}}
\caption{(Color online) Direction of the applied electric field. } \label{fig-smp1}
\end{figure}

In classical electrodynamics, the Laplace equation (i.e.,  $ \nabla^{2}\Phi=0 $ ) in a cylindrical coordinate system has a  general form  given by equation (\ref{2.1}) \cite{Dav99} :
\begin{equation}
\label{2.1}
\frac{\partial^{2}\Phi}{\partial r^{2}} + \dfrac{1}{r}\frac{\partial\Phi}{\partial r} + \dfrac{1}{r^{2}}\frac{\partial\Phi}{\partial \theta^{2}} + \frac{\partial\Phi}{\partial z^{2}}=0.
\end{equation}
Figure~\ref{fig-smp1}  shows that an external electric field $ E_{\text{h}} $ is applied perpendicularly to the axis of a cylindrical inclusion. In a cylindrical coordinate system, the potential is two-dimensional and choosing it is not a function of $z$-axis. In this case,  equation (\ref{2.1}) can be reduced to the form
\begin{equation}
\label{2.2}
\frac{\partial^{2}\Phi}{\partial r^{2}} + \dfrac{1}{r}\frac{\partial\Phi}{\partial r} + \dfrac{1}{r^{2}}\frac{\partial\Phi}{\partial \theta^{2}} =0.
\end{equation}

The solution of equation (\ref{2.2}) can yield the  potential distribution at different regions of a composite of metal coated dielectric core cylindrical nanoinclusions embedded in a linear dielectric host matrix as:
\begin{align}
\Phi_{1}&= -E_{\text{h}}Ar\cos \theta, \qquad r \leqslant r_{1}\,,\\
\Phi_{2}&= -E_{\text{h}}\left(Br - \frac{C}{r}\right)\cos \theta, \qquad r_{1} \leqslant r \leqslant r_{2}\,,\\
\Phi_{3}&= -E_{\text{h}}\left(r - \frac{D}{r}\right)\cos \theta, \qquad r \geqslant r_{2}\,,
\end{align}
where $\Phi_{1} $,  $\Phi_{2} $ and $\Phi_{3} $   are potentials in the dielectric core, metallic shell and the dielectric host matrix, respectively, $ E_{\text{h}} $ is the electric field in the host, $r_{1} $  and $ r_{2}$ are the radii of the dielectric core and the metal shell, respectively, and  $A$, $B$, $C$ and $D$ are unknown coefficients.

The unknown coefficients can be obtained from the electrostatics boundary conditions. Under the long-wave approximation, the case in which the wavelength of the incident electromagnetic wave is greater than the size of the particle, we can use the following electrostatics boundary conditions in order to get the value of the coefficients \cite{Boh83}.

$\bullet$ Potential is continuous at the interface, therefore:
\begin{align}
\label{2.6}
\Phi_{1}&= \Phi_{2}\,, \qquad r=r_{1}\,,\\
\Phi_{2}&= \Phi_{\text{h}}\,, \qquad r=r_{2}\,.
\label{2.7}
\end{align}

$\bullet$ Displacement vector is continuous at the interface
\begin{align}
\label{2.8}
  \epsilon_{1}\frac{\partial\Phi_{1}}{\partial r}&= \epsilon_{2}\frac{\partial\Phi_{2}}{\partial r}\,, \qquad r=r_{1}\,,\\
 \epsilon_{2}\frac{\partial\Phi_{2}}{\partial r}&= \epsilon_{\text{h}}\frac{\partial\Phi_{\text{h}}}{\partial r}\,, \qquad r=r_{2}\,,
 \label{2.9}
\end{align}
where $ \epsilon_{1} $, $ \epsilon_{2}  $ and $  \epsilon_{\text{h}} $ are the dielectric constants of the dielectric core, metal shell and dielectric host, respectively.

Solving equations (\ref{2.6}), (\ref{2.7}),  (\ref{2.8}) and  (\ref{2.9}) simultaneously, can give us the value of the unknown coefficients as listed below:
\begin{align}
\label{2.10}
A&=\frac{4\epsilon_{2}\epsilon_{\text{h}}}{p\Delta}\,,\\
B&=\frac{2\epsilon_{\text{h}}(\epsilon_{1}+ \epsilon_{2})}{p\Delta}\,,\\
C&=\frac{2\epsilon_{\text{h}}(\epsilon_{1}- \epsilon_{2})r_{1}^{2}}{p\Delta}\,,\\
D&=\left\{1-2\epsilon_{\text{h}}\frac{[\epsilon_{2}(2-p)+\epsilon_{1}p]}{p\Delta}\right\}r_{2}^{2}\,,
\end{align}
where $ p=1-(r_{1}/r_{2})^{2} $ is the metal volume fraction in the inclusion, $ \Delta=\epsilon_{2}^{2}+q\epsilon_{2}+\epsilon_{1}\epsilon_{\text{h}} $, and  $ q=(2/p-1)(\epsilon_{1}+\epsilon_{\text{h}}) $.

\section{Resonant frequencies and enhancement factor of local field}
In general, the dielectric constant of the host is complex, but for simplicity we take it real. The dielectric function of the metal in the inclusion is chosen to be in the Drude form \cite{Sis13}
\begin{align}
\epsilon_{2}&=\epsilon_{\infty} - \frac{1}{z(z+\ri\gamma)}\,,\\
\epsilon_{2}'&=\epsilon_{\infty} - \frac{1}{z^{2} +\gamma ^{2}}\,,\\
\epsilon_{2}''&=\frac{\gamma}{z(z^{2} +\gamma ^{2})}\,,
\end{align}
where $\epsilon_{2}' $ and $\epsilon_{2}'' $ are real and imaginary parts of $ \epsilon_{2}$, respectively, $ z=\omega/\omega_{\text{p}} $ is dimensionless frequency, $ \omega $~is the frequency of the incident radiation, $ \omega_{\text{p}} $ is the plasma frequency of the inclusion metal part, $ \nu $ is the electron collision frequency and $ \gamma=\nu/\omega_{\text{p}} $.

The dielectric function of the dielectric core is given by
\begin{equation}
\epsilon_{1}=\epsilon_{10}+\chi\lvert E_{\text{h}}\rvert^{2},
\end{equation}
where  $ \epsilon_{10} $ is the linear part of $\epsilon_{1} $,  $ \chi $ is known as nonlinear Kerr coefficient (nonlinear susceptibility) and for weak field  $ \chi\lvert E_{\text{h}}\rvert^{2} \ll \epsilon_{10}$. At intense incident electromagnetic fields (laser radiation), it is necessary to consider the nonlinear part of the dielectric.

The local field in the core is constant and it can be determined from the gradient of the electrical potential in the core \cite{Sis13}.
\begin{equation}
E_{\text{loc}}= -\nabla \Phi_{1} = A\cdot E_{\text{h}}\,.
\end{equation}

The mathematical expression of the local field enhancement factor for different assemblage of the cylindrical nanoinclusions in terms of the dielectric constants and the metal volume fraction in the inclusion is presented in the following three subsections.
\subsection{Metal coated cylindrical nanoinclusions}

Since $A$ is a complex quantity, it is appropriate to take the square of its module, which is a real quantity. Hence, the expression for the local field enhancement factor $\lvert A\rvert ^{2} $ is as follows:
\begin{equation}
\lvert A\rvert ^{2}= \rho\left(\frac{\beta}{\eta +\mu}\right),
\end{equation}
where
\begin{align}
\rho&=\frac{16\,\epsilon_{\text{h}}^{2}}{p^{2}}\,,\\
\beta&=\epsilon_{2}'^{\,2} + \epsilon_{2}''^{\,2},\\
 \eta &=(\epsilon_{2}'^{\,2}-\epsilon_{2}''^{\,2} + q\epsilon_{2}' + \epsilon_{1}\epsilon_{\text{h}})^{2},\\
 \mu &= \epsilon_{2}''^{\,2}(q+2\epsilon_{2}')^{2}, \\
 q&=\left(\frac{2}{p}-1\right)(\epsilon_{1}+\epsilon_{\text{h}}).
\end{align}
\subsection{Dielectric coated cylindrical nanoinclusions}
In the case of dielectric coated metal cylindrical nanoinclusions, the expression for the local field enhancement factor can be determined from equation (\ref{2.10}) by changing $\epsilon_{1}  $ to $ \epsilon_{2} $ and $\epsilon_{2}  $ to $ \epsilon_{1} $.

\begin{equation}
\lvert A\rvert ^{2}= \rho^{*}\left(\frac{1}{\eta^{*} +\mu^{*}}\right),
\end{equation}
where
\begin{align}
 \rho^{*}&=\frac{16\,\epsilon_{1}^{2}\epsilon_{\text{h}}^{2}}{p^{2}}\,,\\
 \eta^{*}&=(\epsilon_{1}^{2}+ q^{*}\epsilon_{1} + \epsilon_{2}'\epsilon_{\text{h}})^{2},\\
 \mu^{*}&=\left[\left(\frac{2}{p}-1\right)\epsilon_{1} + \epsilon_{\text{h}}\right]^{2}\epsilon_{2}''^{\,2},  \\
 q^{*}&=\left(\frac{2}{p}-1\right)(\epsilon_{2}' + \epsilon_{\text{h}}).
\end{align}
\subsection{Pure metal cylindrical nanoinclusions}
The potential distribution for pure metal cylindrical nanoinclusions having radius $r_{2}  $  in a dielectric host matrix is:
\begin{align}
\Phi_{\text{in}}&= -E_{\text{h}}Ar\cos \theta,  \qquad r \leqslant r_{2}\,,\\
\Phi_{\text{out}}&= -E_{\text{h}}\left(r +\frac{B}{r}\right)\cos \theta, \qquad r > r_{2}\,,
\end{align}
where $\Phi_{\text{in}} $ and $ \Phi_{\text{out}} $ are potentials inside the cylinder and outside the cylinder, respectively, $A$ and $B$ are unknown coefficients which can be obtained from electrostatic boundary conditions.
\begin{align}
A&=\frac{2\epsilon_{\text{h}}}{\epsilon_{2} + \epsilon_{\text{h}}}\,,\\
B&= \left(\frac{\epsilon_{2}-\epsilon_{\text{h}}}{\epsilon_{2}+\epsilon_{\text{h}}}\right)r^{2}_{2}.
\end{align}
Therefore, the local field enhancement factor for a pure metal cylindrical inclusion becomes:
\begin{equation}
\lvert A \rvert ^{2}= \frac{4\,\epsilon_{\text{h}}^{2}}{\left[(\epsilon_{2}' + \epsilon_{\text{h}})^{2} + \epsilon_{2}''^{\,2}\right]}\,.
\end{equation}
\section{Numerical results and discussion}

All the numerical values of the dielectric functions of the composite used in this section are taken from~\cite{Sis13} and table \ref{t1} shows constant values used in numerical calculations.
\begin{table}[!h]
\vspace{-4mm}
\caption{Constant values used in numerical calculations.}
  \begin{center}
    \begin{tabular}{| c | c |}
    \hline \hline
    Constant& Value \\ \hline \hline
    $\epsilon_{1}$ & 6 \\ \hline
            $\epsilon_{\text{h}}$ & 2.25 \\ \hline
      $\omega_{\text{p}}$ & $1.45 \times 10^{16}$  \\  \hline
      $\nu$ & $1.68 \times 10^{14}$ \\ \hline
      $\gamma$ & $1.15 \times 10^{-2}$ \\
    \hline \hline
    \end{tabular}
  \end{center}
  \vspace{-4mm}
  \label{t1}
\end{table}

\subsection{Metal coated dielectric nanoinclusions}

\begin{figure}[!t]
\vspace{-1mm}
\centerline{\includegraphics[width=0.7\textwidth]{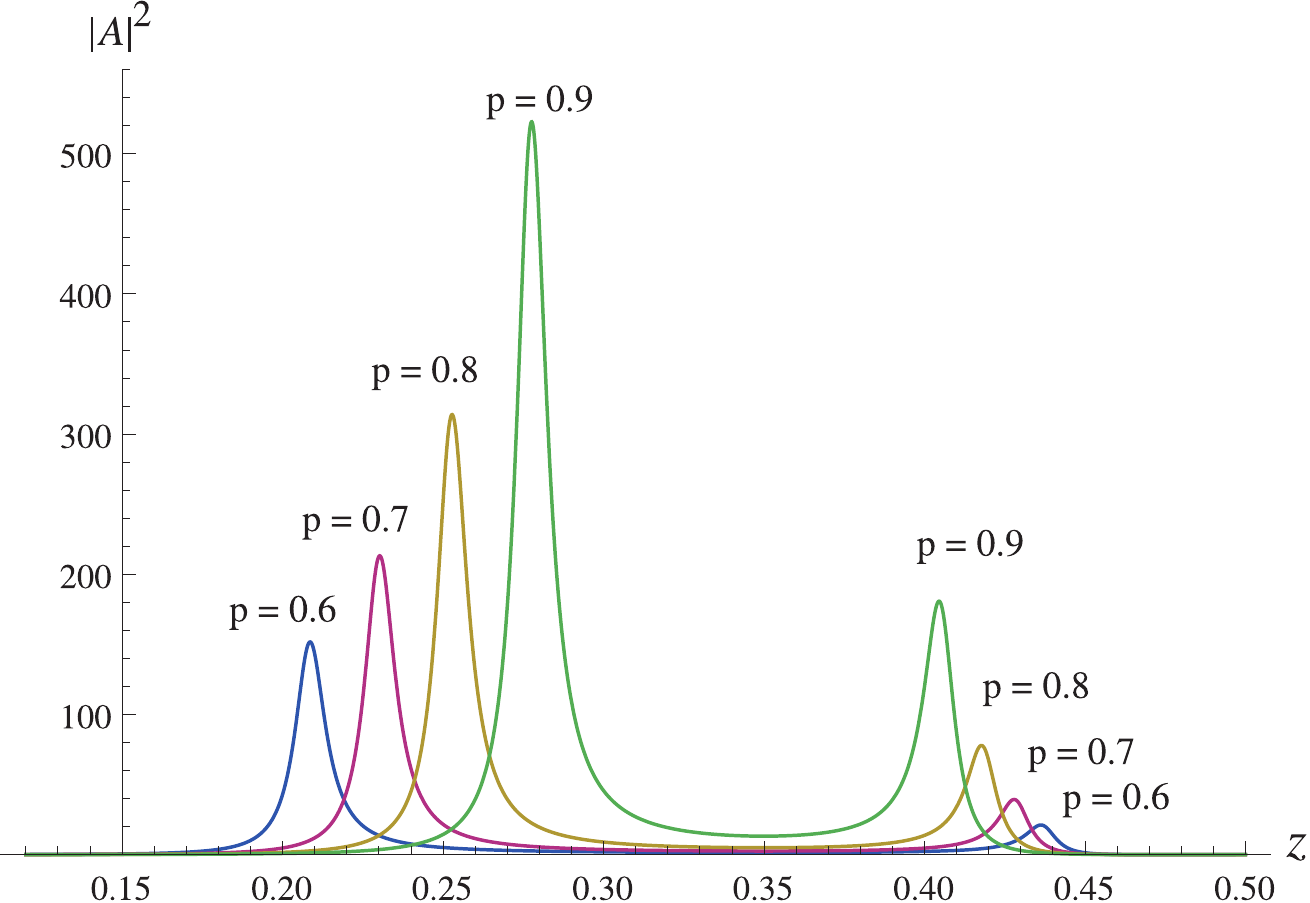}}
\caption{(Color online) Enhancement factor $ |A|^2 $ versus dimensionless frequency $z$ of cylindrical metal inclusion with a dielectric core at different metal fraction in the inclusion $p$; $ \epsilon_{\infty}=4.5. $
} \label{fig-smp2}
\end{figure}

\begin{figure}[!b]
\vspace{-2mm}
\centerline{\includegraphics[width=0.7\textwidth]{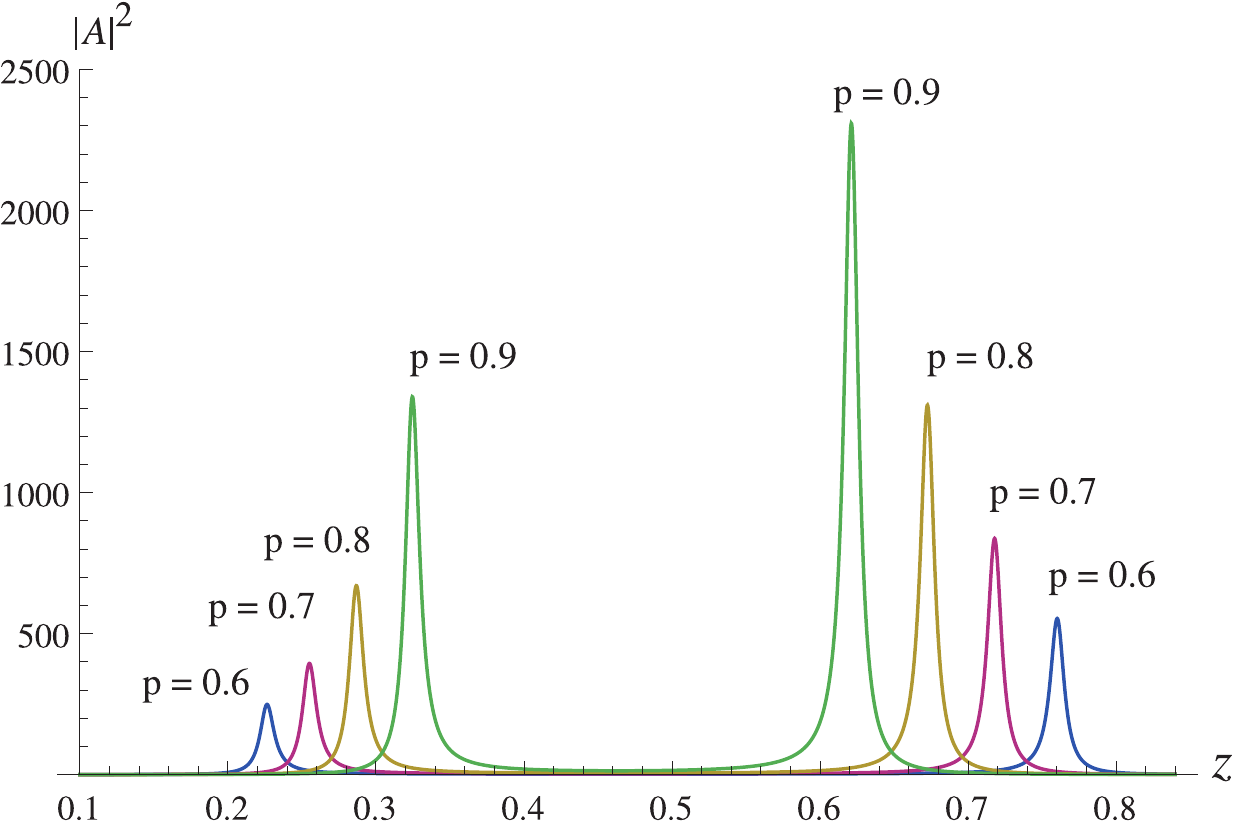}}
\caption{(Color online) Enhancement factor $ |A|^2 $ versus dimensionless frequency $z$ of cylindrical metal inclusion with a dielectric core at different metal fraction in the inclusion $p$; $ \epsilon_{\infty}=1. $
} \label{fig-smp3}
\end{figure}

The series of figures presented below,  $ |A|^2 $ versus $z$, show that the local field enhancement factor varies with $z$. For a composite with metal coated dielectric cylindrical nanoinclusions,  figure~\ref{fig-smp2}  and figure~\ref{fig-smp3}  depict that $ |A|^2 $ has two peak values at two different resonant frequencies. The existence of two resonant frequency is related to the number of interfaces that the metal part of the inclusion has with the two dielectric parts of the composite. The metal shell has  interfaces with the host dielectric and the core dielectric. Hence, the free electrons of the metal oscillate with different surface plasmone frequency along the interface with the dielectric host matrix and along the interface of the core dielectric. At these two resonant frequencies, the enhancement factor of the local field shows a significant increase.


 Sisay and Mal'nev showed that the value of $ \epsilon_{\infty}$ plays a significant role in determining the extent of $ |A|^2 $ for  a composite with spherical inclusions. This study is also capable of confirming the role of $ \epsilon_{\infty}$ in varying the magnitude of $ |A|^2 $ for a composite with cylindrical inclusions. As we decrease the $ \epsilon_{\infty}\,$, the extent of $ |A|^2 $ increases and the second maxima become more important. Based on the obtained numerical results for metal volume fraction $ p = 9 $ and  $ \epsilon_{\infty}=4.5 $, an incident electromagnetic wave can be enhanced around 550 times at the first resonant frequency. For  $ \epsilon_{\infty}=1 $  and $p = 9 $, an incident electromagnetic wave can be enhanced around 2500 times at the second resonant frequency.

 The second quantity which determines the magnitude of $ |A|^2 $ is the volume fraction of the metal in the inclusion. For  $ \epsilon_{\infty}=4.5$ and $ \epsilon_{\infty}=1$, the magnitude of the enhancement factor increases as the metal fraction in the inclusion (i.e., $p$) increases, in addition to the two peak values becoming closer to each other.

\begin{figure}[!t]
\vspace{-1mm}
\centerline{\includegraphics[width=0.7\textwidth]{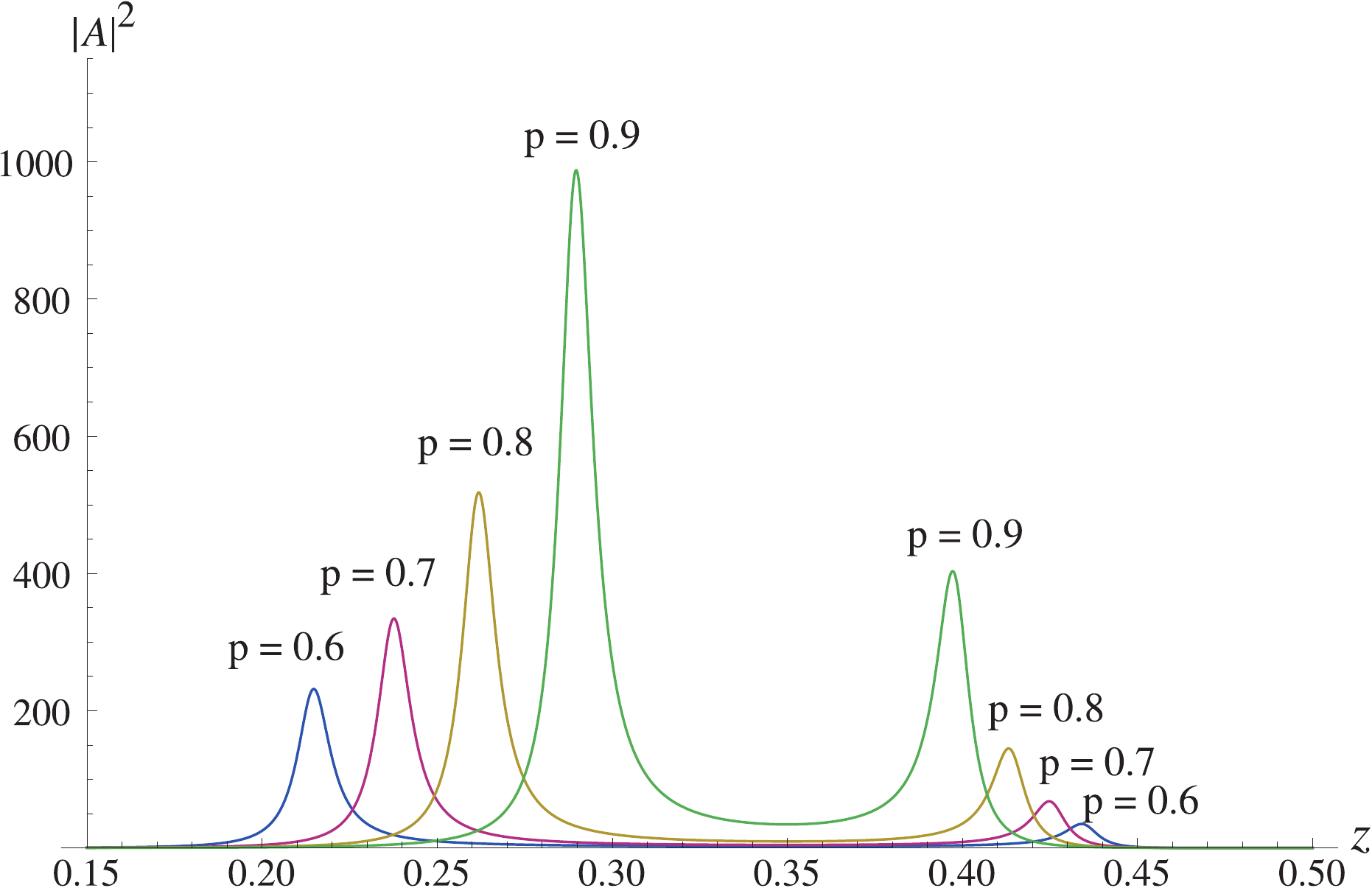}}
\caption{(Color online) Enhancement factor $ |A|^2 $ versus dimensionless frequency $z$ of spherical metal inclusion with a dielectric core at different metal fraction in the inclusion $p$; $ \epsilon_{\infty}=4.5. $
} \label{fig-smp4}
\end{figure}
\begin{figure}[!b]
\vspace{-1mm}
\centerline{\includegraphics[width=0.7\textwidth]{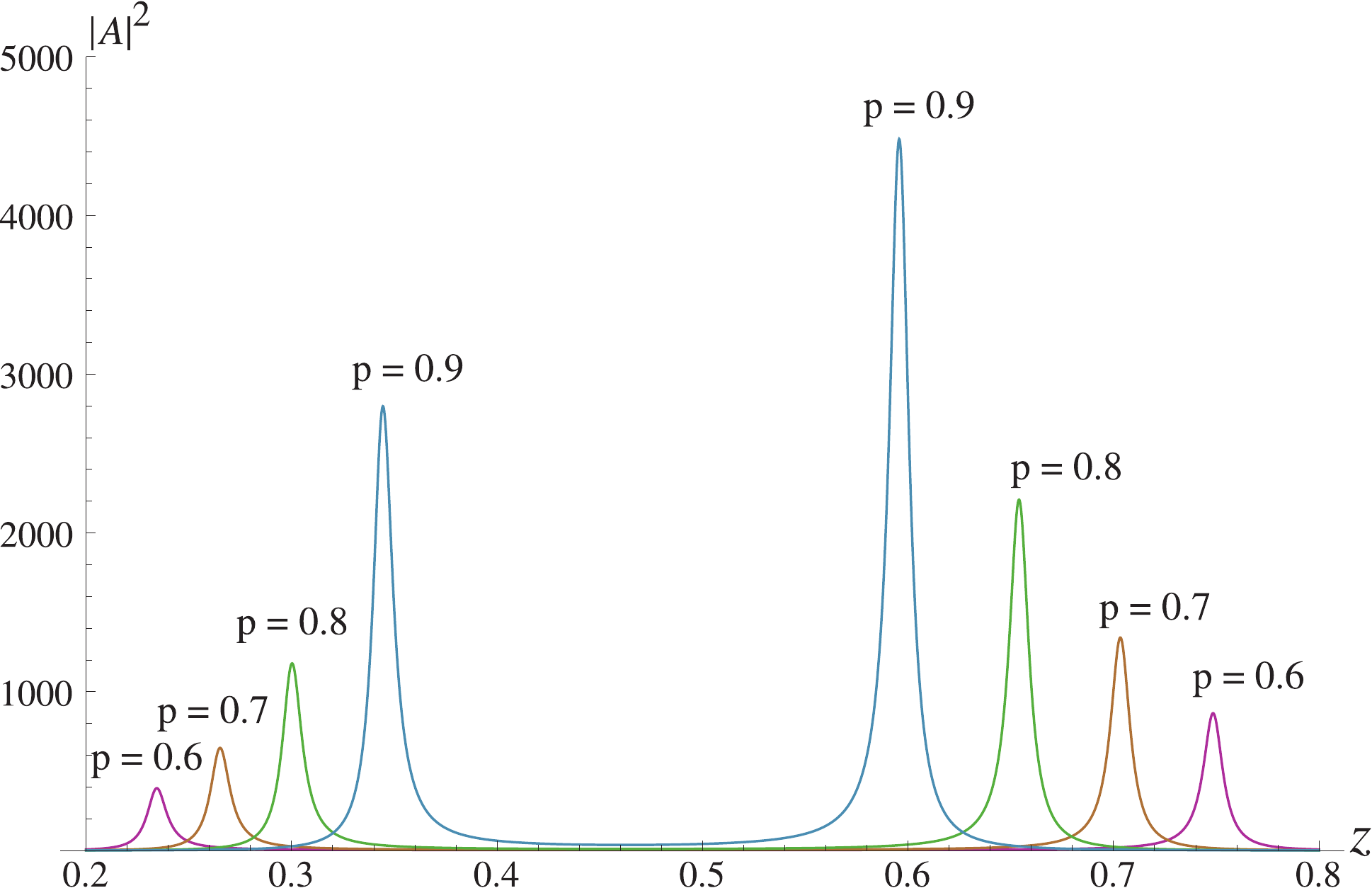}}
\caption{(Color online) Enhancement factor $ |A|^2 $ versus dimensionless frequency $z$ of spherical metal inclusion with a dielectric core at different metal fraction in the inclusion $p$; $ \epsilon_{\infty}=1. $
} \label{fig-smp5}
\end{figure}

In addition to the value of p and $ \epsilon_{\infty}\,$, the geometry of the nanoinclusions can be considered as a factor in obtaining a better enhancement of the local field.  In order to compare the  significant magnitude variation of $ |A|^2 $ due to the change in the geometry of the inclusions, we have presented the results for spherical inclusion in figure~\ref{fig-smp4}  and figure~\ref{fig-smp5}. As Sisay and Mal'nev obtained and the authors of this paper numerically confirmed, for the same $p$, $ \epsilon_{\infty}$ and for other dielectric parameters, the value of $ |A|^2 $ for a composite with spherical inclusion is greater by more than two times when compared with a composite having cylindrical inclusions.

For $ \epsilon_{\infty}=4.5 $  and $p=9$, the enhancement factor for a composite having spherical inclusions has a magnitude of 1000 at the first resonant frequency. The result of our numerical calculation suggests that for the same dielectric parameter and $p$ value, the magnitude of the enhancement factor for the cylindrical inclusion reduces to 500 at the region of the first resonant frequency.

\subsection{Dielectric coated metal cylindrical nanoinclusion}
\vspace{1mm}
\begin{figure}[!t]
\vspace{1mm}
\centerline{\includegraphics[width=0.7\textwidth]{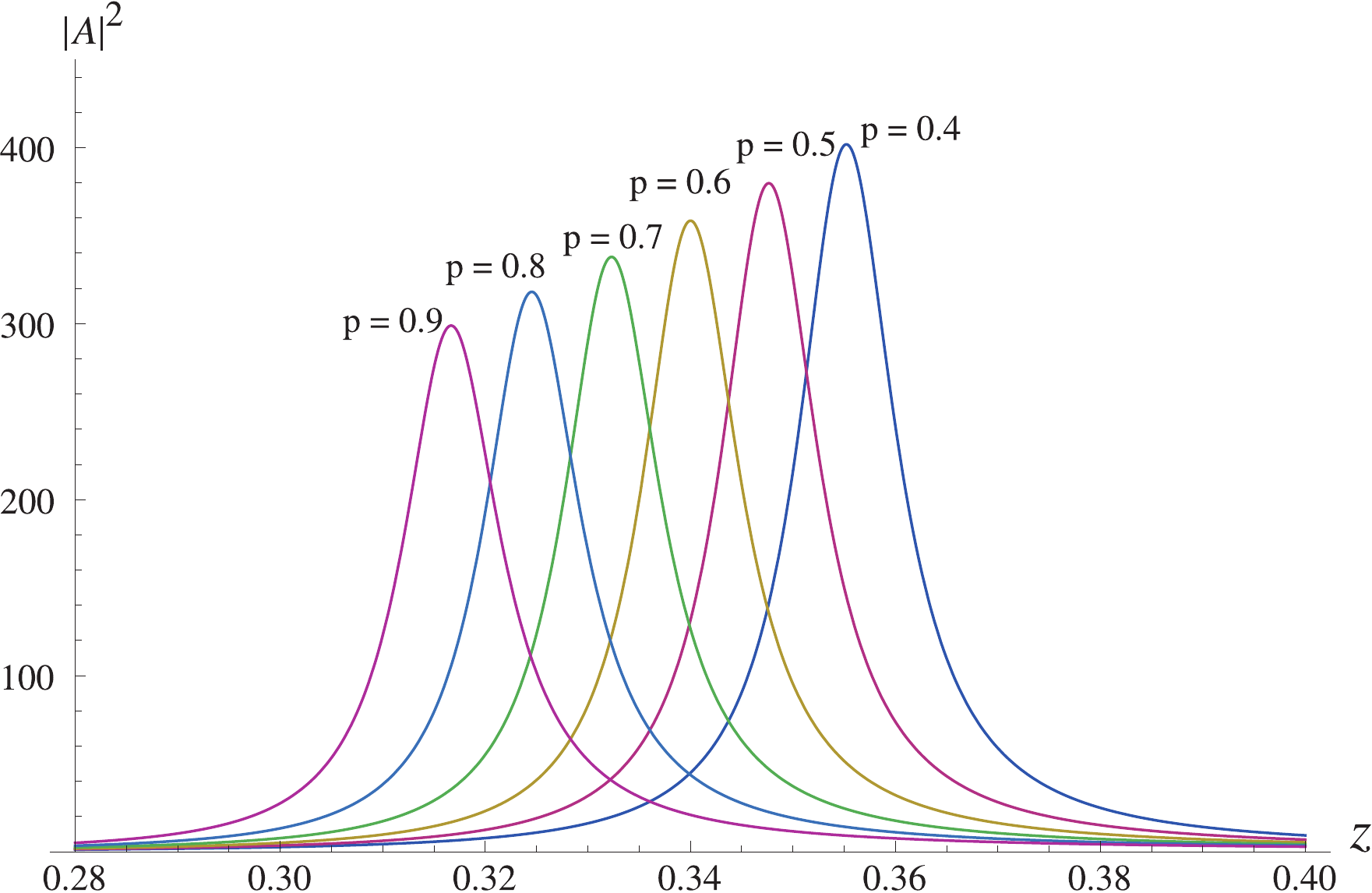}}
\caption{(Color online) Enhancement factor $ |A|^2 $ versus dimensionless frequency $z$ of cylindrical metal inclusion covered by dielectric  at different metal fraction in the inclusion $p$; $\epsilon_{\infty}=4.5$.
} \label{fig-smp6}
\end{figure}
\begin{figure}[!b]
\centerline{\includegraphics[width=0.7\textwidth]{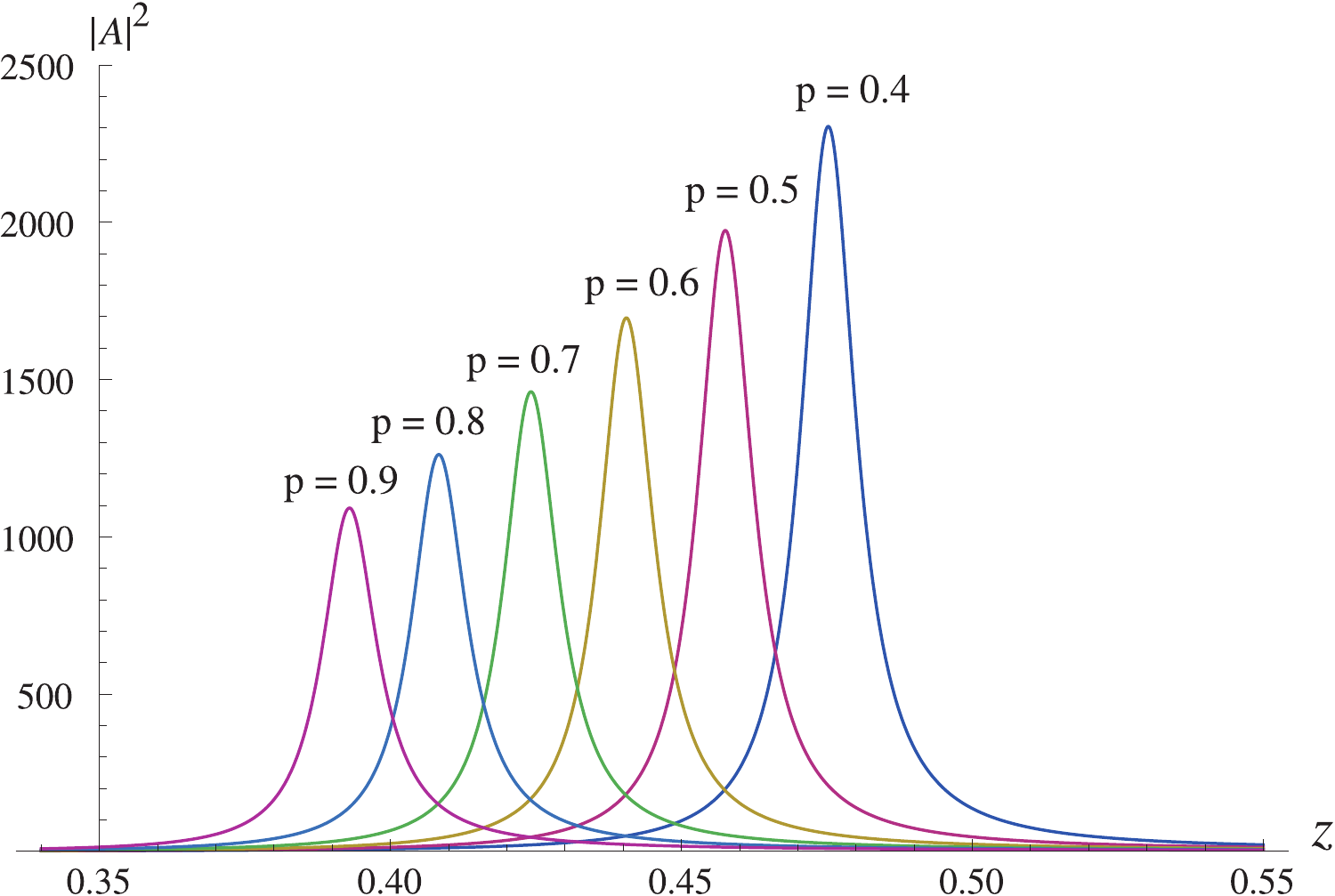}}
\caption{(Color online) Enhancement factor $ |A|^2 $ versus dimensionless frequency $z$ of cylindrical metal inclusion covered by dielectric  at different metal fraction in the inclusion $p$; $\epsilon_{\infty}=1$.
} \label{fig-smp7}
\end{figure}

Figure~\ref{fig-smp6}  and figure~\ref{fig-smp7} shows that for a composite with dielectric coated metal cylindrical nanoinclusions, there is only one resonant frequency and one maximum value of the enhancement factor. For this type of a composite, varying the thickness of the metal plays a role in obtaining a higher value of the enhancement factor. We can easily observe from the results in figure~\ref{fig-smp6}  and figure~\ref{fig-smp7} that, as we decrease the volume fraction of the metal core, the magnitude of the local field enhancement factor rises.

Here again, $ \epsilon_{\infty} $ determines the magnitude of the enhancement factor and $ |A|^2 $ is higher for  $ \epsilon_{\infty} = 1$ than for  $ \epsilon_{\infty}=4.5 $. For $ p=0.9 $ and $\epsilon_{\infty}=4.5$, an incident electromagnetic wave can be enhanced about 400~times at the core of the inclusions. For $ p=0.9 $ and $ \epsilon_{\infty}=1$, the enhancement factor significantly increases in magnitude and it is obtained that the incident field is enhanced beyond 2000.

\subsection{Pure metal cylindrical nanoinclusions}

\begin{figure}[!h]
\vspace{-2mm}
\centerline{\includegraphics[width=0.64\textwidth]{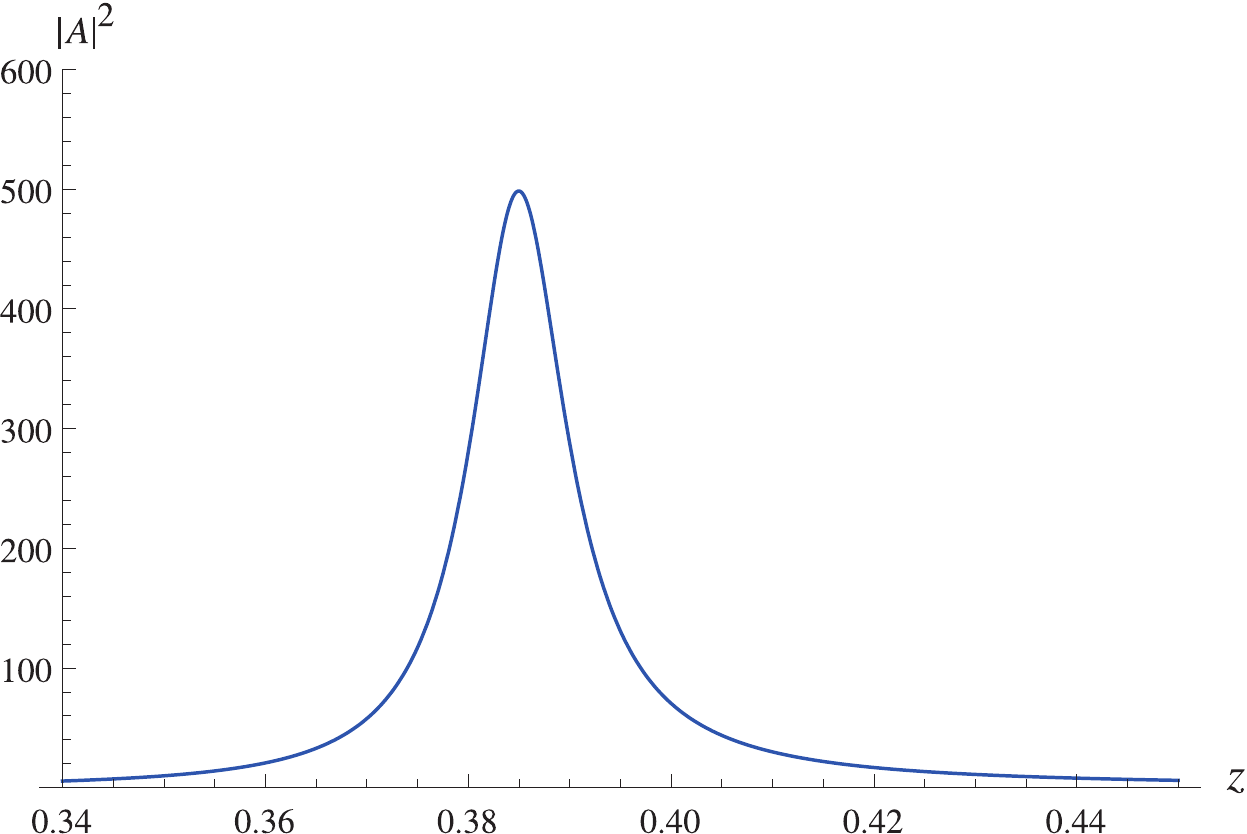}}
\caption{(Color online) Enhancement factor $ |A|^2 $ versus dimensionless frequency $z$ of pure metal  cylindrical inclusion;  $\epsilon_{\infty}=4.5$.
} \label{fig-smp8}
\end{figure}

\begin{figure}[!h]
\vspace{-2mm}
\centerline{
\includegraphics[width=0.64\textwidth]{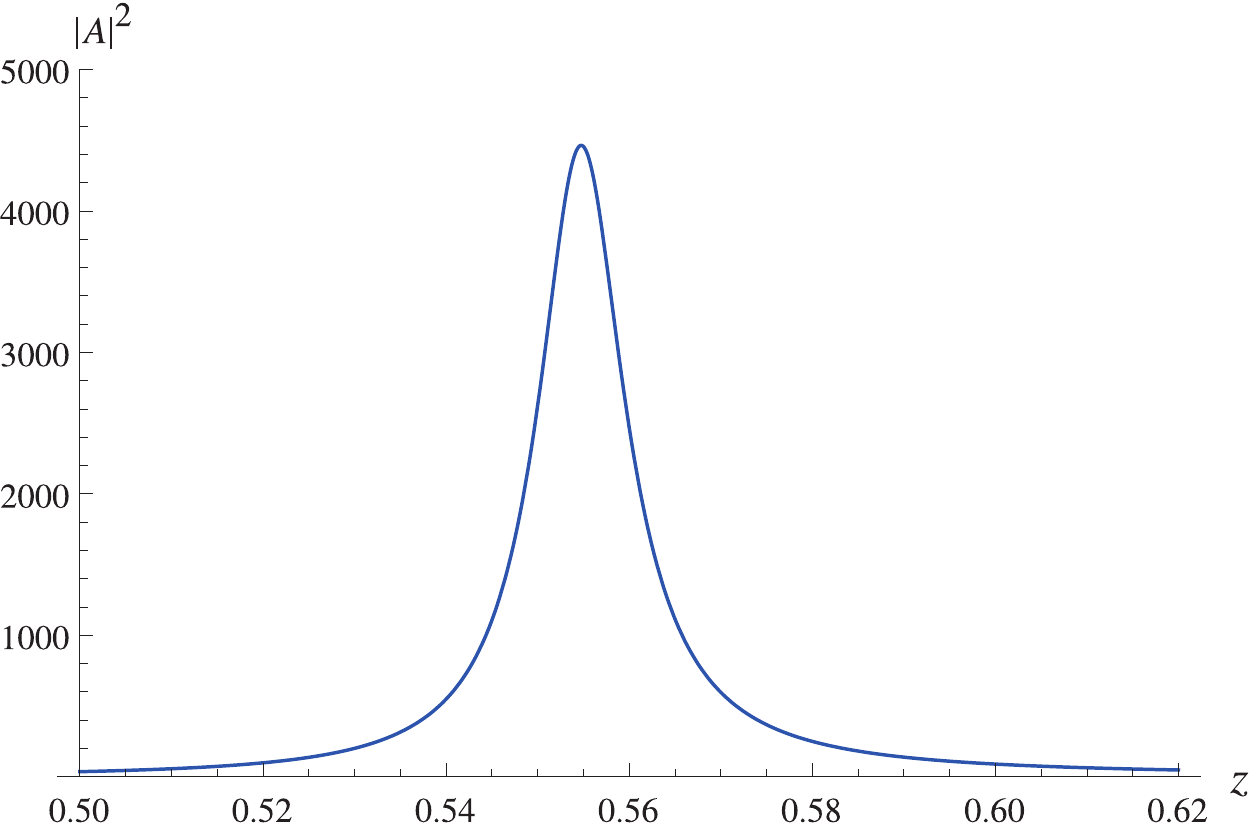}
}
\caption{(Color online) Enhancement factor $ |A|^2 $ versus dimensionless frequency $z$ of pure metal  cylindrical inclusion;  $\epsilon_{\infty}=1$.
} \label{fig-smp9}
\end{figure}

For pure metal inclusions, the main factor that determines the enhancement factor is $\epsilon_{\infty}\,$.
Figure~\ref{fig-smp8}  and figure~\ref{fig-smp9} show that there is only one maximum value of the enhancement factor   at one resonant frequency. The magnitude of the enhancement factor significantly increases as we decrease the value of~$\epsilon_{\infty}\,$. The value of $ |A|^2 $ for $\epsilon_{\infty}=4.5$ is around 500. For $\epsilon_{\infty}=1$, we have obtained a very significant rise in the magnitude of $ |A|^2 $ and numerically it is beyond 4000.

\section{Conclusions}
\vspace{-1mm}

The numerical results we have obtained in our study show that an incident electromagnetic wave can be significantly enhanced at the core of cylindrical nanoinclusions. For a composite with metal coated dielectric nanoinclusions, there are two peak values of the enhancement factor at two different resonant frequencies. The second maxima of the enhancement factor becomes more important for a larger volume of the metal part of the inclusion. For dielectric coated metal and pure metal inclusions,  there is only one maximum of the enhancement factor.

For coated inclusions, the maximum value of the enhancement factor can be determined by varying the thickness of the metal in the inclusion and $\epsilon_{\infty}\,$. For metal coated dielectric inclusions with thick metal cover and small value of $\epsilon_{\infty} $ as well  for dielectric coated metal inclusion with thin metal core and small value of $\epsilon_{\infty}\,$, the enhancement factor becomes more important.

Unlike the coated inclusion, in pure metal inclusion, the only factor that determines the magnitude of the enhancement factor is $\epsilon_{\infty}\,$. Lowering the value of $\epsilon_{\infty} $ significantly raises the magnitude of the enhancement factor. This type of a composite is more preferable and more promising for the occurrence of optical bistability based on its capability of enhancing the incident electromagnetic wave better than the composites with coated nanoinclusions.
\section*{Acknowledgement}

We gratefully acknowledge and dedicate this paper to our former instructor and thesis advisor Professor Vadim N. Mal’nev. May his soul rest in peace!

\ukrainianpart

\title[]%
{Підсилення локального поля  в зоні циліндричних нановключень, вбудованих у лінійну діелектричну   матрицю%
}%
\author[]{Й.А.~Аббо\refaddr{label1,label2}, В.Н. Мальнєв \refaddr{label2}, A.A.~Ісмаїл\refaddr{label2}}
\addresses{
\addr{label1} Аддис Абеба (Аско), P.O.BOX 171078, Ефіопія
\addr{label2} Фізичний факультет, Університет м. Аддис Абеба, P.O.BOX 1176, Аддис Абеба, Ефіопія
}

\makeukrtitle

\begin{abstract}
В цій статті ми обговорили теоретичні концепції, а також представили числові результати підсилення локального поля для різних компонувань
 циліндричних нановключень метал/діелектрик, вбудованих у лінійну діелектричну базисну матрицю. Отримані результати показують, що для композиту з пометалованими включеннями існує два пікових значення коефіцієнта підсилення при  двох різних резонансних частотах. Існування другого максимума стає важливішим для більшої об'ємної частки металічної частини включення. Для металевих включень з діелектричним покриттям є тільки одна резонансна частота і одне пікове значення коефіцієнта підсилення. Підсилення електромагнітної хвилі є обнадійливим для існування нелінійного оптичного явища такого як бістабільність, яке є важливим для оптичного зв'язку і оптичних розрахунків (оптичні комутатори і елементи пам'яті).

\keywords нанокомпозит, циліндричні нановключення, локальне поле, коефіцієнт підсилення, резонансна частота,  оптична бістабільність
\end{abstract}

\end{document}